\documentclass[conference]{IEEEtran}

\IEEEoverridecommandlockouts
\usepackage{cite}
\usepackage{amsmath,amssymb,amsfonts}
\usepackage{graphicx}
\usepackage{textcomp}
\usepackage{xcolor}
\usepackage{multirow}
\usepackage{tabularx}
\usepackage[left=1in, right=1in, top=1in, bottom=1in]{geometry}

\usepackage{caption}
\usepackage{lipsum}  
\def\BibTeX{{\rm B\kern-.05em{\sc i\kern-.025em b}\kern-.08em
    T\kern-.1667em\lower.7ex\hbox{E}\kern-.125emX}}

\usepackage[mediumspace,mediumqspace,Grey,squaren]{SIunits}
\usepackage[bottom]{footmisc}
\usepackage{subcaption} 
\usepackage{makecell}
\usepackage{comment}
\usepackage{enumitem}
\usepackage{float}
\usepackage{color, colortbl}
\definecolor{Gray}{gray}{0.9}
\usepackage{amsmath}
\usepackage[LGR,T1]{fontenc}
\DeclareMathAlphabet{\mathgtt}{LGR}{cmtt}{m}{n}
\usepackage[ruled, vlined]{algorithm2e} 

\begin{document}

\title{FIRETWIN: Digital Twin Advancing Multi-Modal Sensing, Interactive Analytics for Tactical Wildfire Response \thanks{This work is supported by funding from NASA under the award number 80NSSC23K1393, supplement number 80NSSC23K1393-P0001 and the National Science
Foundation under Grant Numbers CNS-2232048, CNS-2038759, and CNS-2204445.}\\
}
\author{ 
Mayamin Hamid Raha$^{1}$, 
Ali Reza Tavakkoli$^{1}$,
Chris Webb$^{2}$,
Mobin Habibpour$^{2}$, \\
Janice Coen$^{3,4}$, 
Eric Rowell$^{5}$,
Fatemeh Afghah$^{2}$
\thanks{$^{1}$Department of Computer Science and Engineering, University of Nevada, Reno, NV, USA,}
\thanks{$^{2}$Holcombe Department of Electrical and Computer Engineering, Clemson University, Clemson, SC, USA,}
\thanks{$^{3}$NSF National Center for Atmospheric Research, Boulder, CO, USA}%
\thanks{$^{4}$University of San Francisco, CA, USA}%
\thanks{$^{5}$ University of Washington, Seattle, WA USA.}%
}
\maketitle
\begin{abstract}

Current wildfire management systems lack integrated virtual environments that combine historical data with immersive digital representations, hindering deep analysis and effective decision making. This paper introduces FIRETWIN, a cyber-physical Digital Twin (DT) designed to bridge complex ecological data and operationally relevant, high-fidelity visualizations for actionable incident response. FIRETWIN generates a dynamic 3D virtual globe that visualizes evolving fire behavior in real time, driven by output from physics-based fire models. The system supports multimodal perspectives—including satellite and drone viewpoints comparable to NOAA's GOES-18 imagery—enabling comprehensive scenario analysis. Users interact with the environment to assess current fire conditions, anticipate progression, and evaluate available resources. Leveraging Google Maps, Unreal Engine, and pregenerated outputs from the CAWFE coupled weather–wildland fire model, we reconstruct the 39,500+ ha spread of the 2014 King Fire in California's Eldorado National Forest. Procedural forest generation and particle-level fire control enable a level of realism and interactivity not possible in field training.
\end{abstract}

\begin{IEEEkeywords}
Digital Twin, Procedural Generation, Particle Systems, Analytical Tool, Multimodal Sensing, Interactive Platform.
\end{IEEEkeywords}
\begin{figure}[!t]
     \centering
     \includegraphics[width=0.43\textwidth,height=6.1cm]{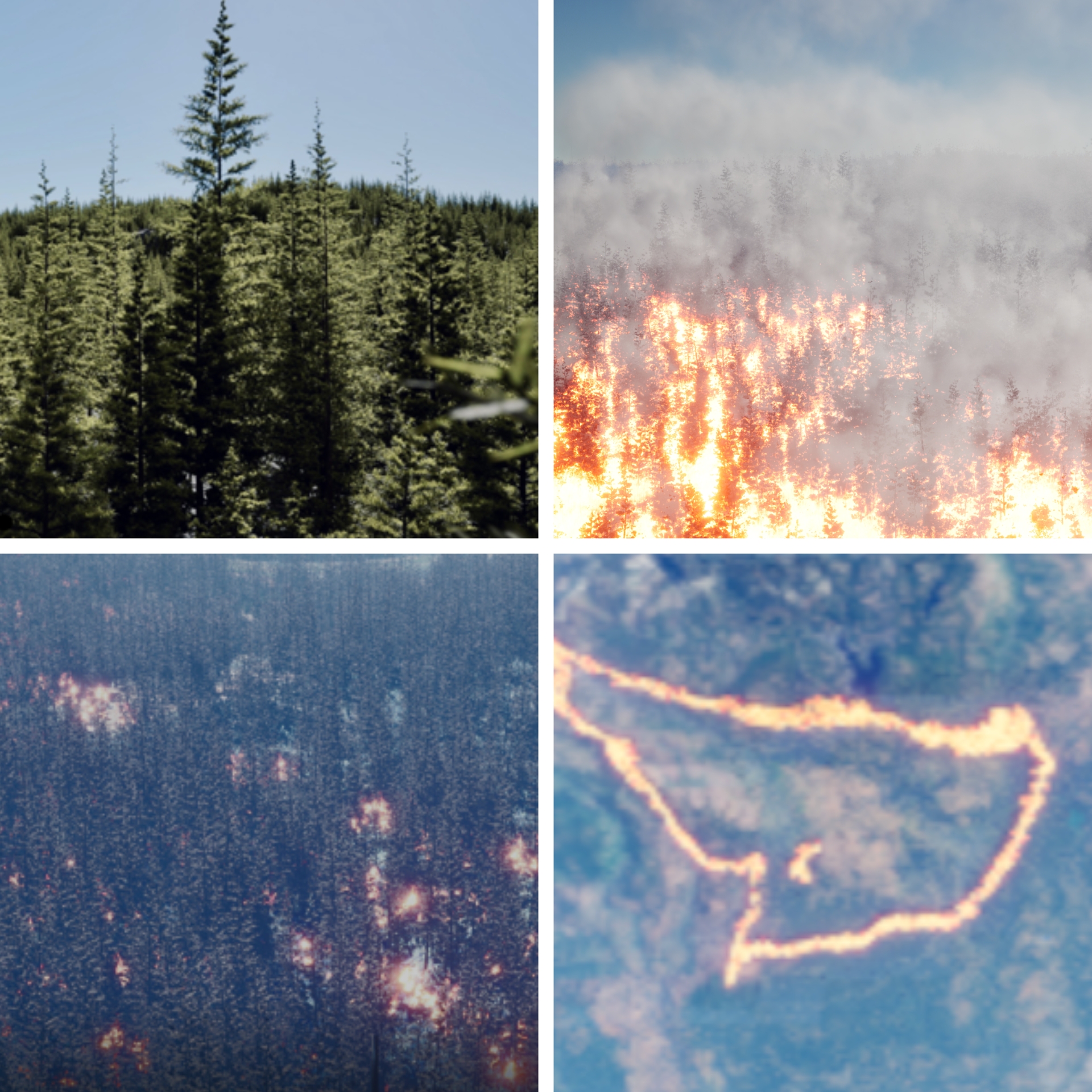}
     \caption{Procedural Forest with High Level of Details (top left), different types of fire spread scenarios (top right, bottom left), and Perimeter of Historic King Fire Reconstruction (bottom right) in FIRETWIN.}
     \label{fig:DT_Fires}
\end{figure}

\section{Introduction}
Gaps in understanding complex fire dynamics and coordinating resource deployment limit effective wildfire management \cite{yun2022novel}. Wildfires have intensified, with burned area increasing 15\% in the past decade, underscoring the need for advanced response tools \cite{guindon2021trends}. Disaster response depends on accurately predicting the location and evolution of an event to ensure timely allocation of resources \cite{Amir_Fire_25}. This challenge requires analytical platforms that use historical data to reveal key patterns in fire spread. Digital twins provide critical disaster management capabilities, including real-time monitoring and spread prediction. However, existing wildfire systems often lack integrated virtual environments that fuse real-time data with immersive visualizations, limiting their effectiveness for accurate analysis and realistic training \cite{yun2022novel}.
A Digital Twin (DT) is a virtual replica of a real-world system, continuously updated by sensor data and physics-based models. Its fidelity depends on the connection between the Physical Twin (PT) and its Cyber Twin (CT) \cite{wu2023comprehensive}. DTs serve as simulation platforms and cost-effective testbeds for exploring what-if scenarios that are difficult to replicate in reality \cite{inyang2025digital}. In wildfire management, DTs offer transformative potential by integrating diverse data into immersive environments that support real-time decision making \cite{li2024review}.
\begin{figure*}[h] 
    \centering
    \includegraphics[width=0.8\textwidth]{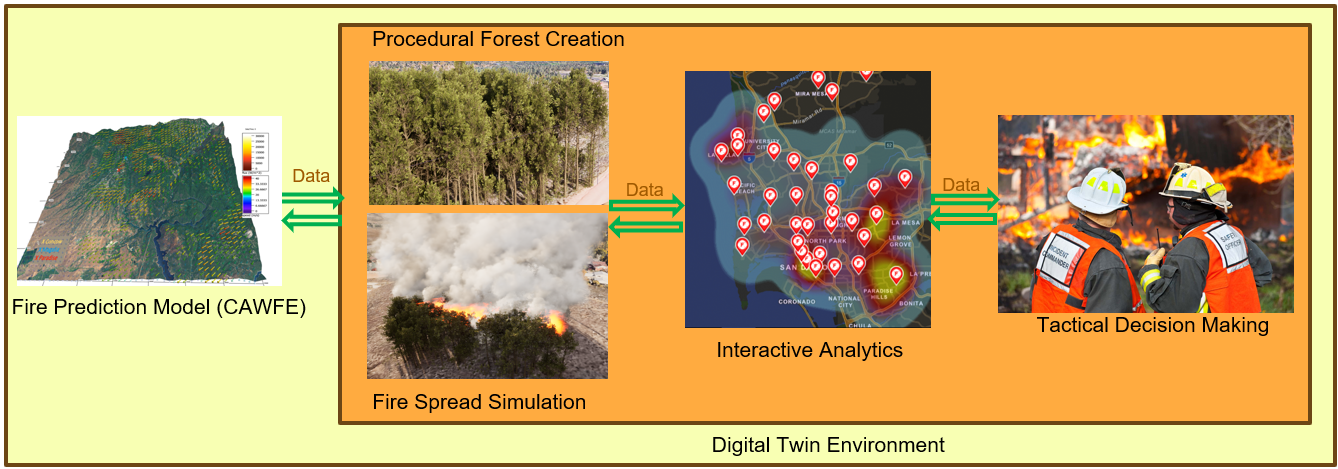}
    \captionof{figure}{High-level data flow within the proposed FIRETWIN environment for wildfire analysis and tactical decision-making.}
    \label{fig:DTCurrentState}
\end{figure*}

Integrating UAVs into a digital twin framework yields a high-fidelity dynamic testbed for autonomous wildfire management. These vital environments enable controlled model validation, realistic training for autonomous algorithms, and synthetic data generation from re-simulated historical fires using archived datasets. Despite these benefits, UAV-integrated fire digital twins are underexplored, presenting a critical opportunity for future research \cite{BOROUJENI2024102369,10644894}.

Current digital twins for wildfires lack 3D physics-based particle-level fire visualizations~\cite{xu2025generative} and immersive environments, which hinders effective situational assessment. They also focus on simple detection and visualization with poor real-time performance~\cite{huang2024modeling}, lacking interactive decision-making tools and multi-modal sensor integration for training. Driven by the need for innovation in wildfire management, we introduce \textit{FIRETWIN}, a comprehensive digital twin framework that offers:
\begin{itemize}[left=0pt]
\item A real-time, geo-synchronized Digital Twin for large-scale geospatial simulation, leveraging Unreal Engine, Google Maps, and Niagara particle systems for wildfire analysis. It integrates high-resolution photogrammetry, terrain, and 3D building models for enhanced realism, utilizing CAWFE model output \cite{coen2024framework} to reconstruct the King Fire \cite{coen2018deconstructing} via physics-backed fire visualization shown in Figure \ref{fig:DT_Fires}.
\item An advanced procedural biome generation system with high-fidelity detail, parameterized vegetation control, and real-world forest data integration to create digital ecosystems at runtime.
\item A proprietary drone flight system with a multimodal sensor suite (RGB, depth, thermal) that integrates a satellite view and minimap for multiperspective wildfire analysis.
\item Interactive analytics from a holistic AI framework, enabling intuitive user interaction, cinematic globe navigation, nearby fire station resource assessment, and an integrated Google Maps reference for tactical decision making.
\item Virtual Reality-enabled global visualization via PC VR integration, offering immersive Digital Twin world exploration with head tracking.
\end{itemize}

 Section II details related work, Section III describes the methodology, Section IV presents results, and Section V offers conclusions and future directions.


\section{Related Work}
Integrating digital twins with wildfire management is a paradigm shift toward data-driven responses that improve situational awareness, resource efficiency, and reduce the impacts of these disasters \cite{cortes2023analysis}. A digital twin is a virtual replica of the real world's objects, systems, and processes. It acts as a tool for simulation and analysis and as an excellent test bed for what-if scenario testing, which can be otherwise very expensive and time consuming to recreate in real life. Visualization of fire model data through Digital Twins is crucial to providing actionable insights to users \cite{li2024review}. Lewis et al. \cite{lewis2024fire} proposed a DT architecture for real-time urban fire tracking and smoke prediction employing 2D Gaussian models and 3D fluid simulations implemented in Blender.  Fan et al. \cite{fan2025digital} develop a digital twin system for predicting forest fire reignition that uses logistic regression with historical data to model smoldering-to-flaming transitions. Ayden et al. \cite{aydin2024employing} introduced a DT-based fire management system that uses a network state forecaster to deploy DTs in IoT networks. Their method enables Digital Twin to simulate network throughput for forest fire warning systems by modeling the topology of the IoT sensor network using Graph Neural Networks (specifically, Graph Attention Networks) and predicting node transmission statuses using LSTM-based forecasting. Several researchers have proposed digital twin approaches to managing wildfires \cite{hyeong2019novel,wang2024rfwnet} using interactive wildfire spread simulators; however, these systems lack the 3D physics-based particle level fire visualizations and scalability in large environments necessary for effective analysis of megafires.

Unmanned aerial vehicles (UAVs) are essential tools for real-time wildfire monitoring \cite{Fatemeh_DDDAS,9381488} and incident management \cite{Flame_paper,Flame2paper,Flame3_paper}, providing critical multi-spectral imaging and situational awareness for emergency response. The development of autonomous UAVs depends on large-scale public datasets, such as the \textit{FLAME} series \cite{Flame1data,Flame2_data,Flame3_data}, which offer synchronized visual and radiometric thermal imagery to train advanced models. These datasets have spurred innovations like estimating pixel-level temperatures from RGB imagery, reducing sensor payloads, and deep learning frameworks like FlameFinder that detect flames through dense smoke \cite{FlameFinder,marinaccio2025seeingheatcolor}. 

To overcome the constraints of existing systems and explore the realm of UAV-integrated fire digital twins, we introduce FIRETWIN, a comprehensive cyber-physical digital twin framework for tactical wildfire response that combines immersive 3D environments with physics-based fire modeling. FIRETWIN integrates multi-modal sensor integration, procedural forest generation, and real-time particle-level fire simulation on a single platform, in contrast to current methods that concentrate on simple detection and display. Unlike existing approaches that employ static fire representations or generic procedural generation \cite{wang2024rfwnet}, our system introduces three algorithmic innovations: real-time geospatial transformation of 2D CAWFE outputs into particle-based 3D visualizations with mathematical flux-to-visual mapping, fuel-density driven procedural forest generation that creates realistic vegetation distributions based on actual fire model data, and dynamic fire optimization with adaptive LOD that scales from individual fire emitters to landscape-scale visualization. We hypothesize that fusing physics-based models, procedural forests, and UAV-simulated sensors in a digital twin improves tactical decision-making over traditional visualization tools.

\section{FIRETWIN Methodology}
In this section, we outline our overall approach and the technical workflow employed in our FIRETWIN for wildfire analysis (see Fig. \ref{fig:DTCurrentState}). We describe the system's architecture, key implementation strategies, and validation procedures.
Harnessing the transformative potential of DT technology, we propose a comprehensive framework that transcends traditional predictive models to tackle complex physical systems. By providing a platform for integrating real-life simulation results with optimization and interactive technologies, our approach aims to aid precise decision-making in critical scenarios, such as wildfire spread management.
\subsection{System Design and Implementation}
Our \textit{FIRETWIN} platform is built on Unreal Engine 5, selected for its advanced rendering and visualization tools \cite{rahmun2022uav}. It leverages the engine's rendering pipeline for high-resolution, physics-backed 3D modeling \cite{tavakkoli2018game} and visualizes the minute-by-minute 2D/3D CAWFE model output \cite{coen2024framework}. The platform's modular architecture, with separate components for rendering, physics, sensors, and interaction, ensures it is upgradeable and supports seamless integration of new features. It also functions as a comprehensive AI framework, enabling bidirectional data exchange to consume AI predictions (e.g., wildfire forecasting) and generate synthetic training data. Designed for environmental monitoring and disaster management, our DT is an interactive platform for simulating and understanding complex geospatial systems.



\subsection{Fire Simulation}
We developed fire emitters to model fire behavior using Unreal Engine 5’s Niagara particle system \cite{visai2024cinematic}, incorporating physics-based interactions such as aerodynamic drag, wind, gravity, velocity, acceleration, and collisions. This framework enables realistic fire spread under dynamic environmental conditions.

To drive fire visualization, we integrate heat flux outputs from the CAWFE model, allowing real-time modulation of fire intensity, spread, and smoke production. As shown in Algorithm \ref{alg:fire_emitter}, a 144×144 grid of flux values is scanned cell by cell. Non-zero flux cells are geolocated (latitude, longitude, elevation), converted to 3D coordinates, and assigned visual parameters. Flux values are normalized and scaled non-linearly to enhance visual contrast, with higher flux producing larger, brighter flames. Particle-based emitters are then spawned at each active location based on the model’s spatial and intensity data \cite{coen2024framework}. Algorithm \ref{alg:fire_emitter} applies CAWFE to fire visualization mapping per Equations~\ref{eq:flux_normalization}-\ref{eq:scale_mapping}. 

To optimize rendering performance, emitter visibility is dynamically managed based on camera position. Our rendering strategy prioritizes nearby fires, applies distance-based level-of-detail (LOD), limits the number of active emitters, and utilizes object pooling to efficiently recycle fire components.


\begin{algorithm}
\caption{Fire Emitter Placement and Scaling}
\label{alg:fire_emitter}
\KwIn{CAWFE simulated heat flux (144×144 grid), geographic coordinates, elevation data}
\KwOut{Positioned and scaled fire emitters}

\For{each grid cell (X, Y) in CAWFE model output}{
    flux\_value $\leftarrow$ GetFluxAtIndex(X, Y)\;
    \If{flux\_value $\leq$ 0}{
        continue 
    }
    
    latitude $\leftarrow$ LatitudeData[Y]\;
    longitude $\leftarrow$ LongitudeData[X]\;
    height $\leftarrow$ ElevationData[Y $\times$ GridSizeX + X]\;
    unreal\_location $\leftarrow$ TransformToUnrealCoordinates(longitude, latitude, height)\;
    
    normalized\_flux $\leftarrow$ Clamp((flux\_value - MinFlux) / (MaxFlux - MinFlux), 0, 1)\;
    curved\_flux $\leftarrow$ Power(normalized\_flux, 1.5) 
    
    scale\_vector $\leftarrow$ Lerp(MinScale(100,100,100), MaxScale(150,150,150), curved\_flux)\;
    color\_scale $\leftarrow$ Lerp(MinColor(100,100,100), MaxColor(500,500,500), curved\_flux)\;
    
    SpawnNiagaraSystem(unreal\_location, scale\_vector, color\_scale)\;
}
\end{algorithm}
\vspace{-.2em}

\subsection{Mathematical Framework and Calibration}
Equations \ref{eq:flux_normalization}, \ref{eq:power_curve}, \ref{eq:scale_mapping} are used to map CAWFE fire intensity data to visual fire parameters in FIRETWIN

\begin{equation}
F_{norm} = \text{clamp}\left(\frac{F_{i,j} - F_{min}}{F_{max} - F_{min}}, 0, 1\right)
\label{eq:flux_normalization}
\end{equation}

where $F_{i,j}$ represents the flux within grid cell $(i, j)$, while $F_{min}$ and $F_{max}$ indicate the minimum and maximum values across the dataset, respectively.

\begin{equation}
F_{curved} = (F_{norm})^{1.5}
\label{eq:power_curve}
\end{equation}

The power curve with exponent 1.5 is used for visual enhancement for low-intensity fires and to prevent oversaturation of high-intensity regions.

\begin{equation}
\mathbf{S} = \mathbf{S}_{min} + (\mathbf{S}_{max} - \mathbf{S}_{min}) \times F_{curved}
\label{eq:scale_mapping}
\end{equation}

where $\mathbf{S} = (S_x, S_y, S_z)$ represents the 3D scale vector.

\textbf{LOD Optimization Framework:}
Our distance-based level of detail (LOD) system use Equations \ref{eq:lod_function} and \ref{eq:particle_scaling} to dynamically adjust the level of detail of fire emitters according to the user's viewing distance within the scene.

\begin{equation}
LOD(d) = \begin{cases}
0 & \text{if } d \leq 1500m \\
1 & \text{if } 1500m < d \leq 3500m \\
2 & \text{if } d > 3500m
\end{cases}
\label{eq:lod_function}
\end{equation}

\begin{equation}
P_{mult}(LOD) = \begin{cases}
1.0 & \text{if } LOD = 0 \\
0.7 & \text{if } LOD = 1 \\
0.4 & \text{if } LOD = 2
\end{cases}
\label{eq:particle_scaling}
\end{equation}

where $P_{mult}$ represents the particle density multiplier, which is used for computational optimization.

\textbf{Algorithm Calibration:} 
For algorithmic calibration, we determined the hyperparameters through iterative testing using the 2014 King Fire CAWFE dataset. We have selected a power curve exponent value of 1.5 to provide fire representation that correlates with flux magnitude data. Tree size thresholds correspond to fuel density distribution and measurements derived from CAWFE.



    
    
        

\subsection{Procedural Generation of Forest}
Procedural generation in Unreal Engine, commonly known as Procedural Content Generation (PCG), is a technique that enables developers to automatically generate extensive game content—such as landscapes, environments, and entire levels—using algorithms and predefined rules. PCG minimizes the need for manual asset placement, enabling the development of diverse and complex virtual worlds. By employing PCG techniques, our system enables the creation of highly detailed and realistic forested landscapes with diverse vegetation types and variable fuel loads. For the procedural generation of forests using fuel mass data integration from CAWFE, we produce Algorithm~\ref{alg:tree_generation}, which loads surface and canopy fuel values from CAWFE and initializes arrays of tree model variants (LargePine, MediumPine, SmallPine). For each grid cell, it extracts canopy\_fuel and surface\_fuel values. When canopy\_fuel $>$ 0.01, it categorizes trees by fuel density: canopy\_fuel $\geq$ 1.6 creates LARGE trees with scale\_mult 1.1-1.4, canopy\_fuel $\geq$ 0.8 generates MEDIUM trees scaled 0.9-1.2, and lower values produce SMALL trees scaled 0.7-1.0. The algorithm transforms grid coordinates to 3D locations using TransformCoords, applies environmental noise, and calculates rotation. Final tree scale combines normalized fuel value (norm\_fuel) interpolated between 6.0-18.0 using Lerp, multiplied by scale\_mult and random variation (0.8-1.2). It randomly selects a model using idx and spawns trees with SpawnTree. When surface\_fuel $>$ 0 and bSpawnGrass is enabled, it places grass instances at ground level, creating a procedural forest that reflects underlying fuel distribution data.

\begin{algorithm}[htbp]
\caption{Fuel-Based Procedural Tree Generation}
\label{alg:tree_generation}
\KwIn{CAWFE fuel data (720×720), tree models, environment params}
\KwOut{Procedurally placed forest with fuel distribution}

LoadFuelData(SurfaceFuelValues, CanopyFuelValues)\;
InitializeTreeModelArrays(LargePine[2], MediumPine[4], SmallPine[4])\;

\For{each cell (X, Y) in FuelGrid}{
    canopy\_fuel $\leftarrow$ CanopyFuelValues[Y $\times$ GridSize + X]\;
    surface\_fuel $\leftarrow$ SurfaceFuelValues[Y $\times$ GridSize + X]\;
    
    \If{canopy\_fuel $>$ 0.01}{
        \If{canopy\_fuel $\geq$ 1.6}{tree\_category $\leftarrow$ LARGE; scale\_mult $\leftarrow$ Random(1.1, 1.4)\;}
        \ElseIf{canopy\_fuel $\geq$ 0.8}{tree\_category $\leftarrow$ MEDIUM; scale\_mult $\leftarrow$ Random(0.9, 1.2)\;}
        \Else{tree\_category $\leftarrow$ SMALL; scale\_mult $\leftarrow$ Random(0.7, 1.0)\;}
        
        location $\leftarrow$ TransformCoords(DenseLon[X], DenseLat[Y])\;
        location $\leftarrow$ ApplyNoise(location, X, Y)\;
        rotation $\leftarrow$ CalculateRotation(X, Y, location)\;
        
        norm\_fuel $\leftarrow$ Clamp(canopy\_fuel / MaxFuel, 0, 1)\;
        scale $\leftarrow$ Lerp(6.0, 18.0, norm\_fuel) $\times$ scale\_mult $\times$ Random(0.8, 1.2)\;
        
        idx $\leftarrow$ Random(0, TreeModels[tree\_category].length - 1)\;
        SpawnTree(TreeModels[tree\_category][idx], location, rotation, scale)\;
    }
    
    \If{surface\_fuel $>$ 0 AND bSpawnGrass}{
        SpawnGrass(location - (0,0,70), Random(0,360), (100,100,100))\;
    }
}
\end{algorithm}

\subsection{Multi-modal Data Integration}
FIRETWIN integrates geospatial processing, sensor simulation, and real-time data processing capabilities to create a comprehensive DT environment for wildfire simulation and analysis.

\textbf{Geospatial Data Processing:} \textit{FIRETWIN} renders the environment using 3D photorealistic tiles with a High Level of Detail (HLOD) architecture, built from multi-source geospatial data including satellite, aerial, and LiDAR sources. To maintain performance, high-resolution tiles are streamed in real-time based on camera proximity using Nanite virtualized geometry. The system supports global navigation via geographic coordinates, dynamic temporal lighting, and Eye Dome Lighting (EDL) for enhanced depth perception, enabling comprehensive wildfire risk assessment without the performance costs of high polygon density.

\textbf{Sensor Simulation:} We simulate a suite of sensors—including RGB, depth, satellite, and thermal—using custom C++ functions, Unreal Engine blueprints, and Cosy's AirSim integration \cite{lidarsim2022jansen}. A thermal visualization is created by mapping averaged CAWFE temperature data onto a thermal color gradient. This gradient is then additively blended with the 2D RGB minimap by adding the thermal colors to the map's normalized pixel intensities. This technique produces procedurally generated heatmaps that realistically emulate UAV thermal imagery while ensuring the underlying visual features of the map are preserved.

\textbf{Real-time Data Fusion:} We utilize asynchronous multi-threading to execute computationally intensive tasks on dedicated worker threads, separate from the main game thread. This parallel processing architecture allows multiple operations, such as processing various data streams, to run simultaneously without compromising rendering performance or user interface responsiveness. This method is key to achieving real-time data fusion without performance degradation.

\subsection {User Interface Design Methodology:} We designed a user-centric interface to facilitate deep engagement with the digital environment. This UI empowers users to instantly teleport to fire-affected regions for on-the-ground analysis, cross-reference locations with Google Maps, and interpret complex CAWFE wildfire data through dynamic 2D and 3D visualizations. Moreover, the interface supports resource management by marking nearby fire stations as points of interest; users can then navigate to these critical locations and access detailed lists of the resources available, particularly at stations servicing high-risk areas. Furthermore, our interactive user interface design also allows for fire asset search based on the area of coverage, budget, tonnage, operational mode, and availability of resources.

\subsection{Virtual Reality}
\textit{FIRETWIN} platform is compatible with Pimax Virtual Reality (VR) headsets (featuring eye tracking and up to 170-degree field of view) and Meta Quest 3 headsets (featuring enhanced mixed reality capabilities), demonstrating broad compatibility across VR platforms. Users can seamlessly access the complete Digital Twin environment, including the reconstructed King Fire progression and procedurally generated Eldorado National Forest, through head-mounted displays with head tracking capabilities.

\begin{table}[htbp]
    \centering
    \caption{Key Performance Metrics Summary}
    \label{tab:performance_summary}
    \begin{tabularx}{\columnwidth}{|>{\raggedright\arraybackslash}X|r|>{\raggedright\arraybackslash}X|}
        \hline
        \textbf{Performance Metric} & \textbf{Value} & \textbf{Result} \\
        \hline
        GPU Memory Usage          & 4.2+ GB        & High-fidelity rendering \\
        Triangles/Frame(Avg)  & 1.16M          & Polygon Count \\
        Draw Calls/Frame          & 3,129          & Optimized pipeline \\
        Rendering Frame Time      & <5.40 ms       & Real-time performance \\
        Fire Emitter Capacity     & 20,736         & Massive fire simulation \\
        Physics Tick Rate         & 24,000 Hz      & Precise simulation \\
        Display Resolution        & 4K@30fps       & Professional quality \\
        \hline
    \end{tabularx}
\end{table}

\subsection{Validation Approach}

We have validated our system via reconstruction of the King Fire event, comparing our simulation results against actual fire progression data to ensure accuracy and realism.

\section{Results}

This section presents the results of our FIRETWIN framework, demonstrating its technical capabilities and performance characteristics in effectively supporting wildfire analysis and decision support. These results validate the scalability of our digital twin.

\subsection{System Performance and Technical Capabilities}
\textbf{System Configuration:} All performance benchmarks are conducted using RTX 4090 GPU (24 GB VRAM), Intel Core i9-14900KF processor, 64GB DDR4-4800 RAM, and NVMe SSD storage. We have used Unreal Engine 5.3 on Windows 11 Pro.

\textbf{Real-time Rendering Performance:} Our FIRETWIN renders million-polygon scenes at 4K resolution (2160p) and 30 FPS, enabling high-fidelity immersive wildfire analysis. It maintains a 24,000 FPS physics tick rate and 3,129 optimized draw calls per frame, demonstrating an efficient rendering pipeline. Comprehensive metrics from our test system are presented in Table \ref{tab:performance_summary}.

\begin{figure}[h!]
     \centering
     \includegraphics[width=0.48\textwidth,height=4.3cm]{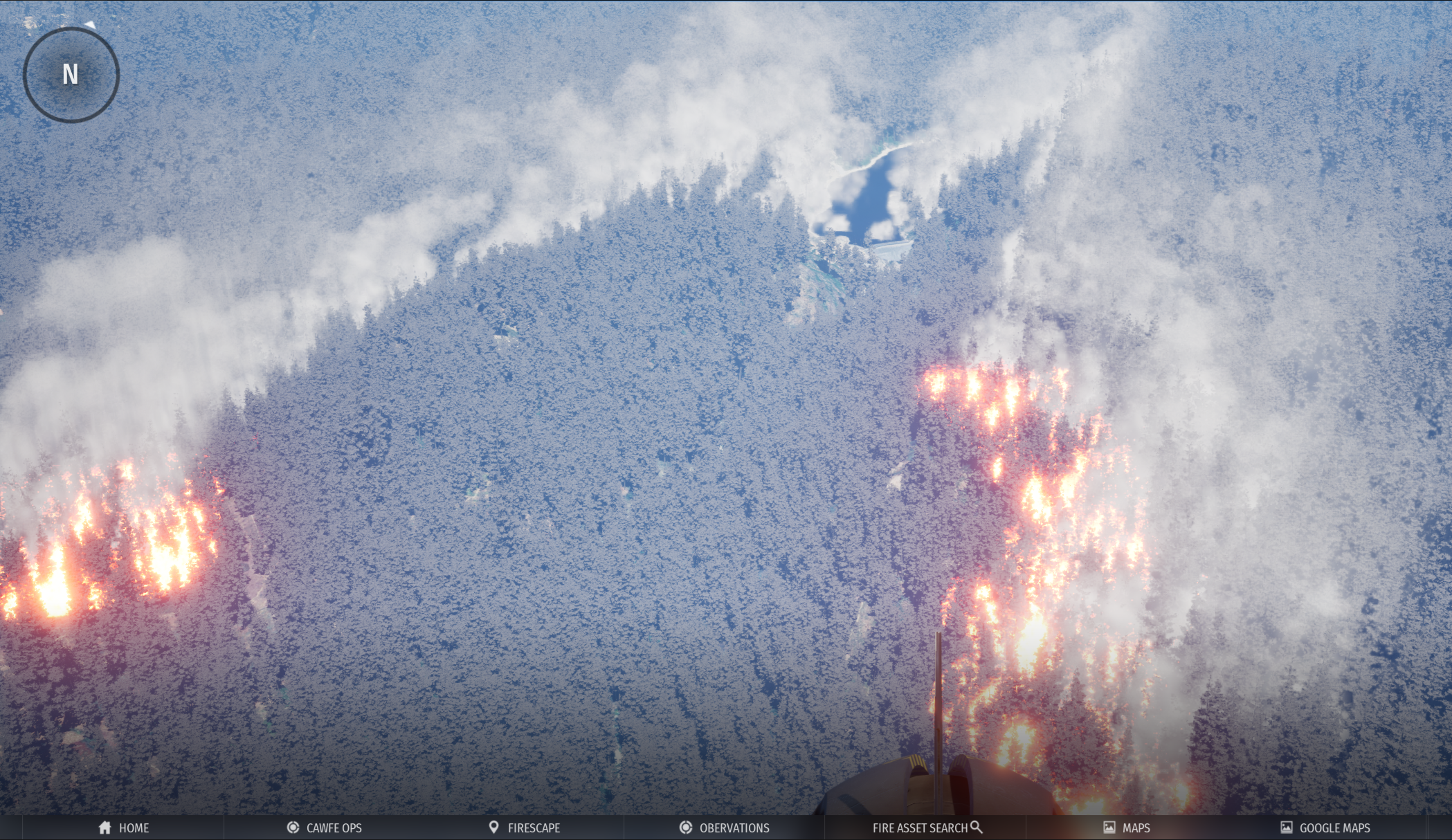}
     \caption{King Fire reconstruction in procedurally generated Eldorado National Forest.}
     \vspace{-.5em}
     \label{fig:kingsFireReconstruction}
\end{figure}

\textbf{Multi-threading Efficiency: } By incorporating asynchronous multi-threading into our design, we can asynchronously load millions of high-resolution Google photorealistic tiles based on viewing distance, simultaneously manage the properties of 20,736 fire emitters, and process CAWFE model output in the background. This approach also enables real-time sensor data fusion in AirSim without blocking the main thread, ensuring wildfire response operations receive the high-fidelity, real-time visualizations that are essential for critical decision-making.

\subsection {Procedural Forest Reconstruction}
We have procedurally constructed Eldorado National Forest in California, representing more than 39,500 hectares of land, using our fuel mass integration algorithm. We spawned 363,632 instances of trees and 419,048 instances of grass for this purpose. In Figure \ref{fig:kingsFireReconstruction}, our procedurally generated forest can be seen from the viewpoint of a drone.

\subsection{Historical Fire Reconstruction}
For evaluation, we integrated fire heat flux output from the CAWFE fire model and reconstructed the 2014 King Fire of California, encompassing more than 39,500 hectares in the Eldorado National Forest (Figure \ref{fig:kingsFireReconstruction}). Our DT is capable of simultaneously spawning 20,736 instances of our fire emitters, where the properties (physics interaction) of each of these fire instances were successfully mapped based on real-world fire model data. In Figure \ref{fig:kingsFireReconstruction}, we can see how the King fire simulation was generated by our geo-synchronized fire DT. The reconstructed fire progression matched historical fire perimeters and intensity patterns.

\begin{figure}[h!]
     \centering
     \includegraphics[width=.43\textwidth]{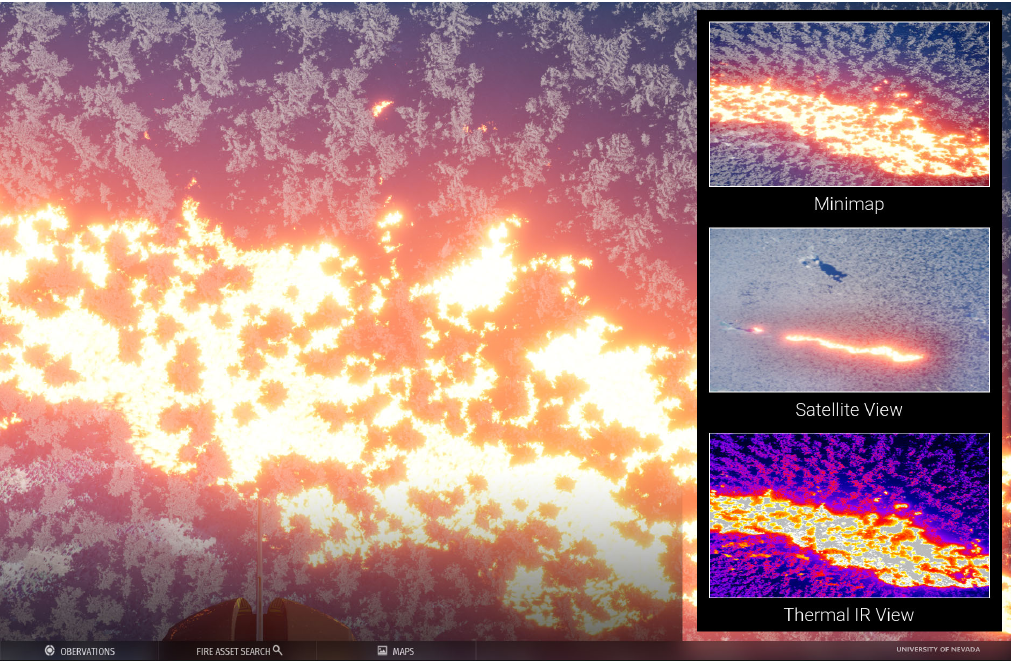}
     \caption{Multi-modal RGB, Satellite and Thermal view Integration in \textit{FIRETWIN} }
     \vspace{-.5em}
     \label{fig:thermal}
\end{figure}

\subsection{Simulated Aerial Sensing}
To complement the broad-scale visualizations discussed previously, our FIRETWIN integrates simulated data from Unmanned Aerial Vehicles (UAVs), which serve as a vital tool for close-range data collection. While satellite imagery is effective for tracking the overall fire perimeter, UAVs provide a different, higher-resolution perspective. This capability is crucial for generating synthetic data for hazardous situations that are difficult or dangerous to monitor in real life, such as fires encroaching on critical infrastructure like power lines or for detailed analysis of prescribed burns where coarse satellite imagery is insufficient.

Our platform generates realistic, multi-modal sensor data from a simulated UAV. This includes high-resolution RGB, depth, and thermal infrared (IR) data streams. To ensure physical realism, the simulated thermal data is derived directly from the temperature outputs of the CAWFE wildfire model. As depicted in Figure \ref{fig:thermal}, these sensor feeds are fused into a unified interface. This allows an operator to view thermal hotspots overlaid with a standard visual map, enabling a direct comparison between fire intensity and the surrounding environmental features. The fusion of these data streams creates a high-fidelity simulation environment, offering a deeper analytical insight for tactical decision-making.

\begin{figure}[h!]
     \centering
     \includegraphics[width=.49\textwidth,height=5.1cm]{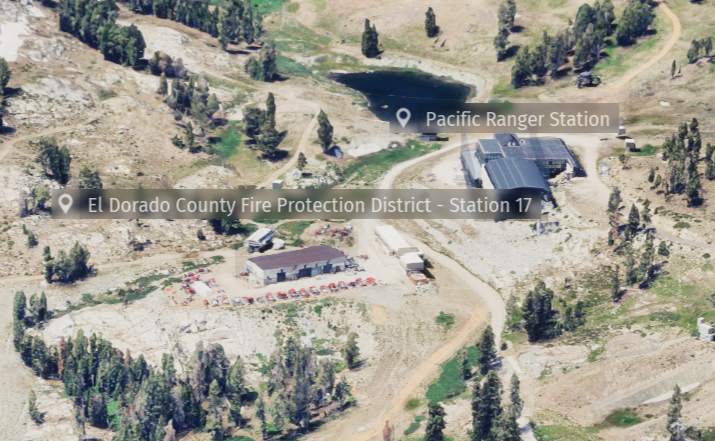}
     \caption{UI featuring point of interest boxes attached to geo-synchronized 3D Firestation models.}
     \vspace{-.5em}
     \label{fig:BP_poi}
\end{figure}

\subsection{Interactive Decision Support Tools}

Our unified interface integrates navigation, visualization and resource assessment tools within a single environment, supporting comprehensive situational awareness and understanding for wildfire response coordination.

\begin{figure}[h!]
     \centering
     \includegraphics[width=.49\textwidth,height=3.6cm]{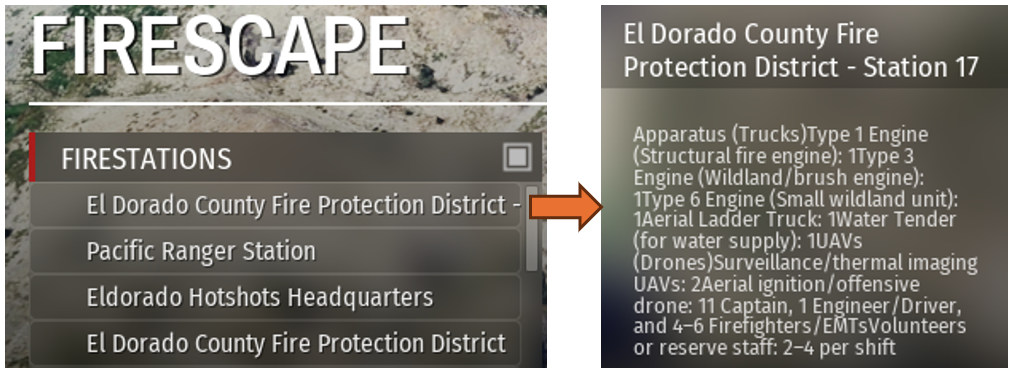}
     \caption{UI featuring firescape functionality with a list of fire stations and enabling user assessment of firestation resources.}
     \vspace{-.5em}
     \label{fig:firescapeDetail}
\end{figure}

\textbf{Navigate to Critical Fire Area:} As shown in Figure \ref{fig:BP_poi}, our design uses point of interest boxes (BP\_POIs) with global anchors for seamless navigation to fire stations near the King Fire via single-button Firescape interactions. BP\_POIs display photographic images of fire stations and detail available firefighting assets (Figure \ref{fig:firescapeDetail}), equipment, and personnel. Additionally, users can instantly teleport to the King Fire's start location for rapid situational assessment. The DT also includes Google Map references with pre-marked stations to provide spatial context between fire perimeters and response resources.

\begin{figure}[h!]
     \centering
     \includegraphics[width=0.45\textwidth,height=4.2cm]{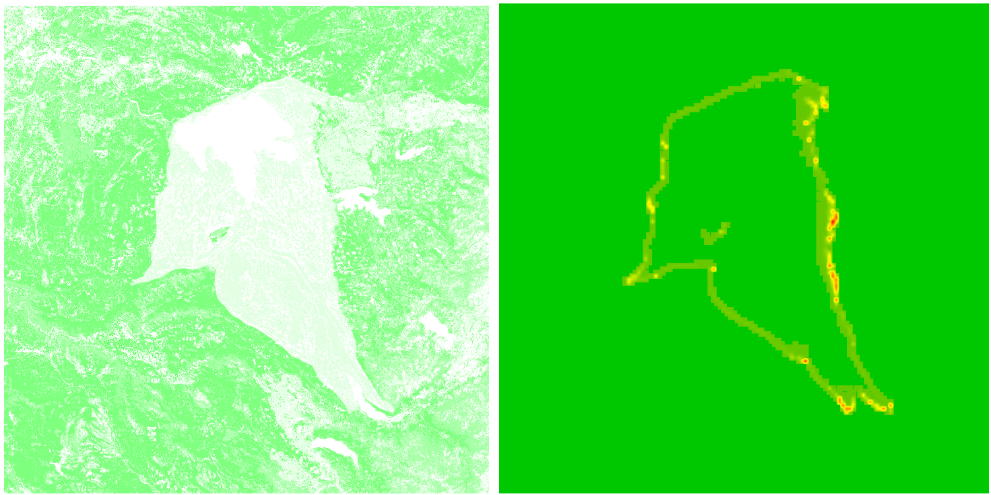}
     \caption{Two-dimensional visualization of Fuel mass and Fire intensity data \cite{coen2024framework}.}
     \vspace{-.5em}
     \label{fig:2Dviz}
\end{figure}

\textbf{Data Visualization and Analysis:}
Our system complements its 3D fire and forest visualizations with 2D data analysis, including fuel mass and fire intensity maps. Figure \ref{fig:2Dviz} shows a fuel heat map created from canopy data \cite{coen2024framework}, where more green indicates denser vegetation (left). On the right, a fire intensity map uses flux values \cite{coen2024framework}, with colors scaling from lighter yellow (low intensity) through deep yellow (medium) to red (high).

\section{Conclusion and Future Scope}
This paper presented a cyber-physical Digital Twin framework that advances tactical wildfire response through dynamic 3D modeling and interactive analytics. By integrating CAWFE fire simulations into Unreal Engine, we developed a platform that enables realistic visualization of fire spread, procedural generation of large-scale environments, and an interactive user interface for decision support. Features such as drone navigation and optimization-based fire emitter rendering further enhance its relevance for wildfire management.
Although the visual fidelity and interactive analytics improve wildfire management, the system requires expensive hardware and operational implementation studies for validation. 
FIRETWIN's potential extends beyond real-time response, supporting applications in operational planning, safety training, and post-incident forensic analysis. Looking ahead, future work will focus on incorporating AI-driven resource allocation models, sensor calibration validation, security protocols, and fault tolerance mechanisms for operational deployment. This will also involve evaluating usability through firefighter training studies and expanding the system with mixed reality capabilities.

\bibliographystyle{IEEEtran}
\bibliography{main}

\begin{thebibliography}{10}
\providecommand{\url}[1]{#1}
\csname url@samestyle\endcsname
\providecommand{\newblock}{\relax}
\providecommand{\bibinfo}[2]{#2}
\providecommand{\BIBentrySTDinterwordspacing}{\spaceskip=0pt\relax}
\providecommand{\BIBentryALTinterwordstretchfactor}{4}
\providecommand{\BIBentryALTinterwordspacing}{\spaceskip=\fontdimen2\font plus
\BIBentryALTinterwordstretchfactor\fontdimen3\font minus \fontdimen4\font\relax}
\providecommand{\BIBforeignlanguage}[2]{{%
\expandafter\ifx\csname l@#1\endcsname\relax
\typeout{** WARNING: IEEEtran.bst: No hyphenation pattern has been}%
\typeout{** loaded for the language `#1'. Using the pattern for}%
\typeout{** the default language instead.}%
\else
\language=\csname l@#1\endcsname
\fi
#2}}
\providecommand{\BIBdecl}{\relax}
\BIBdecl

\bibitem{yun2022novel}
S.-J. Yun, J.-W. Kwon, and W.-T. Kim, ``A novel digital twin architecture with similarity-based hybrid modeling for supporting dependable disaster management systems,'' \emph{Sensors}, vol.~22, no.~13, p. 4774, 2022.

\bibitem{guindon2021trends}
L.~Guindon, S.~Gauthier, F.~Manka, M.-A. Parisien, E.~Whitman, P.~Bernier, A.~Beaudoin, P.~Villemaire, and R.~Skakun, ``Trends in wildfire burn severity across canada, 1985 to 2015,'' \emph{Canadian Journal of Forest Research}, vol.~51, no.~9, pp. 1230--1244, 2021.

\bibitem{Amir_Fire_25}
\BIBentryALTinterwordspacing
S.~Viknesh, A.~Tohidi, F.~Afghah, R.~Stoll, and A.~Arzani, ``Role of flow topology in wind-driven wildfire propagation,'' \emph{Physics of Fluids}, vol.~37, no.~7, p. 076608, 07 2025. [Online]. Available: \url{https://doi.org/10.1063/5.0268416}
\BIBentrySTDinterwordspacing

\bibitem{wu2023comprehensive}
H.~Wu, P.~Ji, H.~Ma, and L.~Xing, ``A comprehensive review of digital twin from the perspective of total process: Data, models, networks and applications,'' \emph{Sensors}, vol.~23, no.~19, p. 8306, 2023.

\bibitem{inyang2025digital}
S.~Inyang and F.~R. Taghikhah, ``Digital twin-based decision support systems for natural disaster management: A systematic review of current trends and approaches,'' \emph{Science Talks}, vol.~13, p. 100406, 2025.

\bibitem{li2024review}
Y.~Li, T.~Zhang, Y.~Ding, R.~Wadhwani, and X.~Huang, ``Review and perspectives of digital twin systems for wildland fire management,'' \emph{Journal of Forestry Research}, vol.~36, no.~1, p.~14, 2024.

\bibitem{BOROUJENI2024102369}
S.~P.~H. Boroujeni, A.~Razi, S.~Khoshdel, F.~Afghah, J.~L. Coen, L.~O’Neill, P.~Fule, A.~Watts, N.-M.~T. Kokolakis, and K.~G. Vamvoudakis, ``A comprehensive survey of research towards ai-enabled unmanned aerial systems in pre-, active-, and post-wildfire management,'' \emph{Information Fusion}, vol. 108, p. 102369, 2024.

\bibitem{10644894}
S.~Khoshdel, Q.~Luo, and F.~Afghah, ``Pyrotrack: Belief-based deep reinforcement learning path planning for aerial wildfire monitoring in partially observable environments,'' in \emph{2024 American Control Conference (ACC)}, 2024, pp. 601--607.

\bibitem{xu2025generative}
H.~Xu, S.~Zlatanova, R.~Liang, and I.~Canbulat, ``Generative ai for predicting 2d and 3d wildfire spread: Beyond physics-based models and traditional deep learning,'' \emph{arXiv preprint arXiv:2506.02485}, 2025.

\bibitem{huang2024modeling}
Y.~Huang, J.~Li, and H.~Zheng, ``Modeling of wildfire digital twin: Research progress in detection, simulation, and prediction techniques,'' \emph{Fire}, vol.~7, no.~11, p. 412, 2024.

\bibitem{coen2024framework}
J.~L. Coen, G.~W. Johnson, J.~S. Romsos, and D.~Saah, ``A framework for conducting and communicating probabilistic wildland fire forecasts,'' \emph{Fire}, vol.~7, no.~7, p. 227, 2024.

\bibitem{coen2018deconstructing}
\BIBentryALTinterwordspacing
J.~L. Coen, E.~N. Stavros, and J.~A. Fites-Kaufman, ``Deconstructing the {K}ing megafire,'' \emph{Ecological Applications}, vol.~28, no.~6, pp. 1565--1580, 2018. [Online]. Available: \url{https://esajournals.onlinelibrary.wiley.com/doi/10.1002/eap.1752}
\BIBentrySTDinterwordspacing

\bibitem{cortes2023analysis}
C.~A.~T. Cortes, S.~Thurow, A.~Ong, J.~J. Sharples, T.~Bednarz, G.~Stevens, and D.~Del~Favero, ``Analysis of wildfire visualization systems for research and training: are they up for the challenge of the current state of wildfires?'' \emph{IEEE transactions on visualization and computer graphics}, vol.~30, no.~7, pp. 4285--4303, 2023.

\bibitem{lewis2024fire}
R.~H. Lewis, J.~Jiao, K.~Seong, A.~Farahi, P.~Navratil, N.~Casebeer, and D.~Niyogi, ``Fire and smoke digital twin--a computational framework for modeling fire incident outcomes,'' \emph{Computers, Environment and Urban Systems}, vol. 110, p. 102093, 2024.

\bibitem{fan2025digital}
W.~Fan, W.~Zai, and W.~Li, ``A digital twin approach to forest fire re-ignition: Mechanisms, prediction, and suppression visualization.'' \emph{Forests (19994907)}, vol.~16, no.~3, 2025.

\bibitem{aydin2024employing}
B.~Aydin and S.~F. Oktug, ``Employing digital twin to forest fire management systems,'' in \emph{2024 9th International Conference on Computer Science and Engineering (UBMK)}.\hskip 1em plus 0.5em minus 0.4em\relax IEEE, 2024, pp. 1--6.

\bibitem{hyeong2019novel}
K.~Hyeong-su, K.~Jin-Woo, S.~Yun, and W.-T. Kim, ``A novel wildfire digital-twin framework using interactive wildfire spread simulator,'' in \emph{2019 Eleventh International Conference on Ubiquitous and Future Networks (ICUFN)}.\hskip 1em plus 0.5em minus 0.4em\relax IEEE, 2019, pp. 636--638.

\bibitem{wang2024rfwnet}
G.~Wang, H.~Li, S.~Ye, H.~Zhao, H.~Ding, and S.~Xie, ``Rfwnet: A multiscale remote sensing forest wildfire detection network with digital twinning, adaptive spatial aggregation, and dynamic sparse features,'' \emph{IEEE Transactions on Geoscience and Remote Sensing}, vol.~62, pp. 1--23, 2024.

\bibitem{Fatemeh_DDDAS}
F.~Afghah, ``Autonomous unmanned aerial vehicle systems in wildfire detection and management-challenges and opportunities,'' in \emph{Dynamic Data Driven Applications Systems}, E.~Blasch, F.~Darema, and A.~Aved, Eds.\hskip 1em plus 0.5em minus 0.4em\relax Cham: Springer Nature Switzerland, 2024, pp. 386--394.

\bibitem{9381488}
A.~Shamsoshoara, F.~Afghah, E.~Blasch, J.~Ashdown, and M.~Bennis, ``Uav-assisted communication in remote disaster areas using imitation learning,'' \emph{IEEE Open Journal of the Communications Society}, vol.~2, pp. 738--753, 2021.

\bibitem{Flame_paper}
\BIBentryALTinterwordspacing
A.~Shamsoshoara, F.~Afghah, A.~Razi, L.~Zheng, P.~Z. FulÃ©, and E.~Blasch, ``Aerial imagery pile burn detection using deep learning: The flame dataset,'' \emph{Computer Networks}, vol. 193, p. 108001, 2021. [Online]. Available: \url{https://www.sciencedirect.com/science/article/pii/S1389128621001201}
\BIBentrySTDinterwordspacing

\bibitem{Flame2paper}
X.~Chen, B.~Hopkins, H.~Wang, L.~O’Neill, F.~Afghah, A.~Razi, P.~Fulé, J.~Coen, E.~Rowell, and A.~Watts, ``Wildland fire detection and monitoring using a drone-collected rgb/ir image dataset,'' \emph{IEEE Access}, vol.~10, pp. 121\,301--121\,317, 2022.

\bibitem{Flame3_paper}
\BIBentryALTinterwordspacing
B.~Hopkins, L.~ONeill, M.~Marinaccio, E.~Rowell, R.~Parsons, S.~Flanary, I.~Nazim, C.~Seielstad, and F.~Afghah, ``Flame 3 dataset: Unleashing the power of radiometric thermal uav imagery for wildfire management,'' 2024. [Online]. Available: \url{https://arxiv.org/abs/2412.02831}
\BIBentrySTDinterwordspacing

\bibitem{Flame1data}
\BIBentryALTinterwordspacing
A.~Shamsoshoara, F.~Afghah, A.~Razi, L.~Zheng, P.~Fulé, and E.~Blasch, ``The flame dataset: Aerial imagery pile burn detection using drones (uavs),'' 2020. [Online]. Available: \url{https://dx.doi.org/10.21227/qad6-r683}
\BIBentrySTDinterwordspacing

\bibitem{Flame2_data}
B.~Hopkins, L.~O'Neill, F.~Afghah, A.~Razi, E.~Rowell, A.~Watts, P.~Fule, and J.~Coen, ``Flame 2: Fire detection and modeling: Aerial multi-spectral image dataset,'' \emph{IEEE DataPort}, 2023.

\bibitem{Flame3_data}
\BIBentryALTinterwordspacing
B.~Hopkins, L.~O'Neill, M.~Marinaccio, F.~Afghah, E.~Rowell, R.~Parsons, and S.~Flanary, ``Flame 3 - radiometric thermal uav imagery for wildfire management,'' 2024. [Online]. Available: \url{https://dx.doi.org/10.21227/w0mz-aq48}
\BIBentrySTDinterwordspacing

\bibitem{FlameFinder}
H.~Rajoli, S.~Khoshdel, F.~Afghah, and X.~Ma, ``Flamefinder: Illuminating obscured fire through smoke with attentive deep metric learning,'' \emph{IEEE Transactions on Geoscience and Remote Sensing}, vol.~62, pp. 1--12, 2024.

\bibitem{marinaccio2025seeingheatcolor}
\BIBentryALTinterwordspacing
M.~Marinaccio and F.~Afghah, ``Seeing heat with color -- rgb-only wildfire temperature inference from sam-guided multimodal distillation using radiometric ground truth,'' 2025. [Online]. Available: \url{https://arxiv.org/abs/2505.01638}
\BIBentrySTDinterwordspacing

\bibitem{rahmun2022uav}
M.~Rahmun, T.~Deb, S.~A. Bijoy, and M.~H. Raha, ``Uav-crowd: Violent and non-violent crowd activity simulator from the perspective of uav,'' \emph{arXiv preprint arXiv:2208.06702}, 2022.

\bibitem{tavakkoli2018game}
A.~Tavakkoli, \emph{Game Development and Simulation with Unreal Technology}.\hskip 1em plus 0.5em minus 0.4em\relax CRC Press, 2018.

\bibitem{visai2024cinematic}
G.~Visai, \emph{Cinematic Photoreal Environments in Unreal Engine 5: Create captivating worlds and unleash the power of cinematic tools without coding}.\hskip 1em plus 0.5em minus 0.4em\relax Packt Publishing Ltd, 2024.

\bibitem{lidarsim2022jansen}
W.~Jansen, N.~Huebel, and J.~Steckel, ``Physical lidar simulation in real-time engine,'' in \emph{2022 IEEE Sensors}, 2022, pp. 1--4.

\end{thebibliography}

\end{document}